\RequirePackage{fix-cm}
\documentclass[twocolumn]{svjour3}          
\smartqed  
\usepackage{graphicx}
\usepackage{booktabs}
\usepackage{caption}
\usepackage{cite}
\usepackage{microtype}
\usepackage[hyphens]{url}
\usepackage{hyperref}
\hypersetup{breaklinks=true}
\usepackage{csquotes} 

\begin{document}
\title{Investigating the effects of Goodreads' challenges on individuals’ reading habits
}


\author{Yasaman Jafari*         \and
        Nazanin Sabri \and
        Behnam Bahrak 
}


\institute{Y. Jafari \at
              School of Electrical and Computer Engineering, University of Tehran \\
              \email{ys.jafari@ut.ac.ir}           
           \and
           N. Sabri \at
              School of Electrical and Computer Engineering, University of Tehran \\
              \email{nazanin.sabri@ut.ac.ir}
            \and
              B. Bahrak \at
              School of Electrical and Computer Engineering, University of Tehran \\
              \email{bahrak@ut.ac.ir}
}

\date{Received: date / Accepted: date}

\maketitle

\begin{abstract}
Sharing our goals with others and setting public challenges for ourselves is a topic that has been the center of many discussions. This study examines reading challenges, how participation in them has changed throughout the years, and how they influence users' reading productivity. To do so, we analyze Goodreads, a social book cataloging website, with a yearly challenge feature. We further show that gender is a significant factor in how successful individuals are in their challenges. Additionally, we investigate the association between participation in reading challenges and the number of books people read.
\end{abstract}

\section{Introduction}
\label{sec:introduction}
Setting goals for yourself in different aspects of your life and publicly announcing those goals is nothing new. All year long, our social media feeds are filled with various instances of month, or year-long fitness \cite{site:fitness_challenge,ehrlen2020shared}, coding \cite{site:coding_challenge}, or art \cite{site:art_challenge} challenges, where users announce they are going to be doing a certain activity for a duration of time or share their progress. Sharing of new years resolutions is also a common occurrence these days \cite{site:instagram_new_years}. Hashtags such as \#paintingchallenge and \#codingchallenge, each have tens of thousands of posts on Instagram, showing many partake in such public advertisements of their goals and progress. But does the setting and announcement of goals on public platforms help or hurt your actual progress?\\
The media and numerous individuals \cite{site:should_you_talk_about_goals,site:dont_talk_about_goals,site:talk_about_your_goals,sivers2010keep} have weighed in with their thoughts on the effects of sharing your goals with others. With some supporting the act and some being strictly against it. There are also many scientific studies on the matter, which we will thoroughly review in Section \ref{sec:related_work}. In this study, we will be focusing on a specific instance of this question: reading goals. And to this end, we will be analyzing Goodreads.\\
Goodreads is a social media platform for readers. Users can form social connections and add their readings on the website. One feature of the website, which is of great interest to this study, is "\textit{Goodreads Reading Challenge}". The challenges are yearly events where users can set a certain number of books they plan to read that year (known as "\textit{pledging}" on the platform) and monitor their progress throughout the year. Goodreads makes it a point for users to see their challenges by showing your progress on your homepage, as well as adding it to your profile. Additionally, users can view the pledged counts and advances of other participants.\\
In this study, we aim to take a closer look at these challenges and answer the following questions:

\begin{itemize}
    \item How has challenge participation changed throughout the years? Have pledged counts and success rates changed since the feature was first introduced?
    \item How do demographic variables influence reading habits?
    \item Does challenge participation (and the public commitment to reading a certain number of books) increase the number of books read?
\end{itemize}
We further extend the study by looking at discussion of the topic on more mainstream social networks, such as Instagram and Twitter, to see how users feel about these challenges and what they share with people outside their reading community.\\
The rest of this paper is structured as follows: In Section \ref{sec:related_work}, a brief overview of related studies is presented. Next, our data collection methods and dataset statistics are discussed in Section \ref{sec:data}. Our results are presented in Section \ref{sec:results}, and finally Section \ref{sec:conclusion} concludes the paper. 
\section{Related Work}
\label{sec:related_work} 
We will begin this section by first providing a review of studies conducted on Goodreads. Then we will briefly review studies on goal setting and sharing and the effects they could have on performance.\\
As Goodreads is primarily a platform for users to add and talk about the books they have read, a large proportion of the body of work on Goodreads is on the analysis of book reviews. In \cite{kousha2017goodreads}, the viability of using Goodreads as a means to assess book impact is put to the test. With a focus on academic books, the impact factor and Goodreads engagement of the books were compared. The authors report that the engagements on the website, while prone to manipulations, could be used to assess books' impact. Additionally, book reading behavior on Goodreads has been shown to predict how well the book will do regarding sales \cite{maity2017book}. The content of reviews are also extensively studied \cite{driscoll2019faraway,parksepp2019sentiment,reisler2019cognitive}. \cite{hajibayova2019investigation} investigates the lingual features of reviews, demonstrating that user-generated reviews are unreliable as they are mostly positive and attempt to persuade others to read the book as well. \cite{shahsavari2020automated} use aggregate reviews as a means for story-graph creation and find their method to be quite accurate compared to the book's true network.\\
Moreover, some studies have looked at specific instances and events concerning Goodreads. For instance, its acquisition by Amazon \cite{albrechtslund2017negotiating}, or some policy changes made by the platform \cite{matthews2016professionals}. \\
Other studies have also investigated the social aspect of Goodreads by examining how users behave on the platform and who they form friendships with \cite{nakamura2013words,thelwall2017goodreads,thelwall2019reader}. \cite{thelwall2017goodreads} finds that men and woman mostly have similar behaviors, with women usually adding more books and usually rating them less favorably. Another study reports significant gender differences in rating books of different genres, finding that users usually rate books by authors of their own gender more favorably \cite{thelwall2019reader}. \cite{sabri2020cross} is another study that takes advantage of the website's users' multinational nature to investigate cross-country reading preferences and the factors that influence these preferences. However, to the best of our knowledge, no studies have been conducted on Goodreads challenges.\\
Goals are defined as "\textit{Desired states that people seek to obtain, maintain, or avoid}" \cite{emmons1989personal,klein2008goal}. Goal-setting theory and goal-choice have been studied extensively, but most prominently in the context of organizations and work. Research has found that more challenging and specific goals result in higher levels of performance \cite{lunenburg2011goal,locke2006new}. Goal-choice is significantly affected by gender, and self-esteem \cite{levy1991effects}. \\
Sharing your goals with others is often viewed through a couple of different lenses. The first is premature praise and the intention-behavior gap. Praise is defined as "\textit{positive evaluations made by a person of another's products, performances, or attributes, where the evaluator presumes the validity of the standards on which the evaluation is based}" \cite{kanouse1981semantics,delin1994praise}. \cite{haimovitz2011effects} explores how person and process praise could affect performance and reports that more generally, process praise increases motivation while personal praise decreases it. \cite{gollwitzer2009intentions} similarly finds that identity-related behavioral intentions that were noticed by others would result in less intense actions. However, since reading is not an identity-related behavior, the finding of \cite{gollwitzer2009intentions} might not apply, and the sharing of progress could potentially result in process praise, which could help improve performance.\\ 
Another view is that of accountability. Accountability is defined as "\textit{stewardship with responsibility for creation and use of resources and a public reckoning of how they are used}" \cite{hubbell2007quality}. A thorough review of the literature on accountability is available in \cite{lerner1999accounting}. While some studies have found accountability to help performance, others show that it is not always the case. To better perform appears to be connected to who you decide to share your goals with \cite{klein2020goals,site:share_but_be_careful_with_who}. Reporting that the group with whom you share your goal must be perceived by you to have a higher status, for the sharing to be effective \cite{klein2020goals}. \\
Social attention is another matter to consider. Research has shown that people act differently when they know they could be observed compared to when they are alone \cite{herman2003effects,kurzban2007audience}. In more detail, studies have demonstrated that performance of simple tasks is improved (in terms of speed and accuracy) in the presence of an audience \cite{zajonc1965social}, while the performance of complex tasks are worsened \cite{bond1983social}. \cite{steinmetz2017beyond} provides an in-depth review of studies on the topic. So whether we consider reading to be a simple or a complex task, the effects based on this theory would be different. 
\section{Data Collection and Preparation}
\label{sec:data}
In this section, the data collection process is first described, then the steps we took to clean and extract various features from the data to prepare it for further analysis are explained.
\subsection{Challenges}
Annual Goodreads challenges are one of the features that help the Goodreads community define a goal, specifying how many books they want to read in the following year. This number is known as the "\textit{pledged}" number of books. These challenges begin every January and finish when the year comes to an end. Users can keep track of the number of books they read during this time, and every book they read will get them closer to their goal. The Goodreads reading challenge data and the associated books are the main datasets used in this research to help us understand how reading challenges affect users' reading habits. The exact features available in this dataset are depicted in Table \ref{tab:challenge}. The data is collected through Goodreads' public API \footnote{https://www.goodreads.com/api}. This dataset includes 5,523,896 instances of challenge data for 4,363,093 unique users and 289,078 books associated with these challenges. We query the API with random challenge identifiers to retrieve this data. Since identifiers are assigned incrementally, with higher numbers corresponding to newer years, we make sure to request numbers distantly apart. With this method, at least 25,549 challenge entries were retrieved for each year from 2011 to 2020. At the time of writing this paper, the users' challenges in 2020 were still in progress, and we decided not to use their information, thus omitted them from the dataset. Moreover, we excluded the information of challenges with more than 500 pledged books or more than 200 read books, as they tend to be outliers. Also, as Goodreads users are members of a reading community, it is logical for them to read at least one book while participating in a challenge; therefore, we did not consider challenges with 0 read books. One possible explanation is that users with 0 read books did not know that they had to record the finish time of the corresponding books or update their read books as Goodreads is not as popular as other social platforms such as Instagram or Twitter. 2,233,517 out of 5,523,896 entries were deleted for this reason; therefore, 3,254,382 challenges for 2,251,574 users remained. The information regarding the number of read and pledged books during these challenges can be seen in Table \ref{tab:challenge_pledged_read}. This information indicates that users tend to overestimate their abilities in reading books. \\
In addition to the data explained above, we collected the data from users' profiles indicating all the challenges they have participated in, including the number of pledged books and read books during each one. This data was then used in order to compare users' performance while participating in a yearly challenge versus the years they were not part of an annual challenge. This dataset has 10,649 entries belonging to 4,558 unique users; the users were selected randomly from the aforementioned pool of users. This data also contained 283 challenges with 0 read books, which were deleted due to the same reason mentioned above. \\
\begin{table*}[htbp]
  \centering
  \captionsetup{justification=centering}
  \caption{Reading Challenge Information}
  \label{tab:challenge}
  \begin{tabular}{c c}
    \hline
    Feature&Description\\
    \hline
    Challenge ID&Unique identifier for a challenge \\
    \hline
    User ID&Unique identifier for the user associated with this challenge \\
    \hline
    Read Count&Number of books this user completely read in this challenge \\
    \hline
    Pledged Count&Number of books this user planned to read in this challenge \\
    \hline
    Year & The year this challenge took place \\
    \hline
    Book IDs& A list of unique identifiers for the books read in this challenge\\
  \hline
\end{tabular}
\end{table*}

\begin{table}[htbp]
  \centering
  \captionsetup{justification=centering}
  \caption{Reading Challenge Pledged and Read Counts}
  \label{tab:challenge_pledged_read}
  \begin{tabular}{c c c}
    \hline
    &Pledged&Read\\
    \hline
    mean&36.59&23.30\\ \hline
    median&25.0&13.0\\ \hline
    standard deviation&34.13&29.25\\
  \hline
\end{tabular}
\end{table}

\subsection{Users and Books}
To analyze demographic features and how such features affect users' participation and reading habits, users' personal information is needed. For this purpose, Goodreads' public API was used again. In order to use the API, user IDs were extracted from corresponding challenges and used to retrieve the aforementioned data. Moreover, books' information, such as their format, is needed for further analysis. This information can also be retrieved similarly. User information and book information dataset columns are shown in Table \ref{tab:user} and Table \ref{tab:book}, respectively.
In total, 35,695 instances of user information were retrieved in Users dataset, among which, only 186 users' \textit{gender}, 1,977 users' \textit{age}, and 5,735 users' \textit{location} were initially available. This should not be confused with the Challenges dataset. It was previously mentioned that 3,254,382 instances of challenge data (after deleting challenges with 0 read books) for 2,251,574 unique users were retrieved, but we do not have these user's personal information. We know that these challenges belonged to 2,251,574 unique users based on their \textit{user ID}. As mentioned, only a small number of users' \textit{gender}, \textit{age}, and \textit{location} were initially available; therefore, these features for the rest of these 35,695 users were extracted by analyzing their profile pictures and personal details. Particularly, for extracting \textit{countries}, both \textit{location} and \textit{about} features were analyzed and the user's country, city, or state name were retrieved in case of availability using \textit{geotext}\footnote{https://pypi.org/project/geotext/} and \textit{pycountry}\footnote{https://pypi.org/project/pycountry/} libraries. This data was then checked manually, and the required corrections were made. Gender detection was done using each user's \textit{name}, \textit{about}, and \textit{image} columns. Moreover, age was detected from the \textit{about} column by finding age keywords such as \textit{age}, \textit{years}, \textit{y/o}, \textit{year} and then categorized in ranges. After doing so, users with ages less than 9 and more than 100 were deleted from the dataset as they are unlikely to be real values. In total, 10\% of users' \textit{age} was detected. Others had not mentioned anything regarding their age. \\
There is also one more dataset that shows which users have read which books, regardless of their participation in reading challenges. This data is then used to compare users' reading habits while they are not participating in any challenges as opposed to the time they are. The columns for this dataset are depicted in Table \ref{tab:userBook}. 

\begin{table}[htbp]
  \centering
  \captionsetup{justification=centering}
  \caption{User Information}
  \label{tab:user}
  \begin{tabular}{c c}
    \hline
    Column&Description\\
    \hline
    User ID&Unique Identifier for a User \\ \hline
    Name&Name of this user \\ \hline
    Image&Profile picture of this user \\ \hline
    Location&User's location - can be country, city, or state \\ \hline
    Gender&Specified as male, female, or unknown \\ \hline
    Age&This is between 9 to 110 \\ \hline
    About&User's "\textit{about me}" information 
    \\
  \hline
\end{tabular}
\end{table}

\begin{table*}[htbp]
  \centering
  \captionsetup{justification=centering}
  \caption{Book Information}
  \label{tab:book}
  \begin{tabular}{c c}
    \hline
    Column&Description\\
    \hline
    Book ID & Unique identifier for a book\\ \hline
    Name & Name of this book\\ \hline
    Format & Format which can be Hardcover, Paperback, Audio, etc.\\ \hline
    Number of Pages & This book's page count\\
  \hline
\end{tabular}
\end{table*}

\begin{table}[htbp]
  \centering
  \captionsetup{justification=centering}
  \caption{User-Book Information}
  \label{tab:userBook}
  \begin{tabular}{c c}
    \hline
    Column&Description\\
    \hline
    User ID & Unique identifier for a user\\ \hline
    Book ID & Unique identifier for a book this user read \\ \hline
    Read At & Date the book was marked as read\\ \hline
    Read Count & Number of times the user read the book \\
  \hline
\end{tabular}
\end{table}

\subsection{Twitter and Instagram}
To analyze how people talk about their reading challenges outside their reading community, we collected Instagram posts and tweets which contained \#readingChallenge and \#goodreadsChallenge hashtags. These hashtags were used in different years, and therefore, the posts belonged to various years. For Instagram, specifically, we also searched for hashtags such as \#2015goodreadsChallenge and \#2015readingChallenge, which included the years. This was not done for Twitter as we had enough data for each year. Furthermore, since our analysis was made on these posts' distributions through months, we did not necessarily need data for all years. In total, 418,913 Instagram posts and 48,320 tweets were collected.  By manually studying these posts and tweets, we realized that they contained book reviews and challenge updates. Using the timestamp of each entry, the month they were posted was extracted, and using the text or caption, the sentiment for each was retrieved. For extracting the sentiments, \textit{TextBlob} library \footnote{https://textblob.readthedocs.io/en/dev/} was used. Texts with sentiment between -0.1 and 0.1 were considered neutral, whereas sentiments between -1 and -0.1 and between 0.1 and 1 were considered negative and positive, respectively. Several examples for the tweets and posts' captions are shown in Table \ref{tab:sentiment_example}. This data was then used to analyze the posts' distribution through the months of the year. The result of which has been reported in the following sections.

\begin{table*}[htbp]
  \centering
  \captionsetup{justification=centering}
  \caption{Posts and Tweets' Sentiment Examples}
  \label{tab:sentiment_example}
  \begin{tabular}{c c}
    \hline
    Text&Sentiment\\
    \hline
    I woke up with a cold!! Oh no!!! At least it got me out of work so now I can work on the& -0.65\\
     \#YASavesChallenge \#readingchallenge \\ \hline
    October is the perfect month to read a book that takes place at night! Check out our recommendations & 0.75 \\
    to fulfill this reading challenge prompt. Our picks are seasonally appropriate with witches \\and vampires! \#booklist \#readingchallenge \#bookclubbelles\\ \hline
    I have read and reviewed 6 books from my TBR list. I am woefully behind schedule when it comes & -0.5\\
     to meeting my Goodreads 2020 challenge, but I'm alright with it because these reads were fantast-\\
    ic! \#ReadingChallenge2020 \#readersofinstagram \#romancereadersofinstagram \#wonderfulbook\\ \#AmReadingRomance \#slowreader \#amreading \#readingromance \\ \hline
    The Perfect Girlfriend" is a pretty good read so far. I usually only say this if I'm 50 pages & 0.45 \\
     in but at 31 pages it's got a pace set. So far, so good. \#readingtime \#readingchallenge2020 \#bookworm \\
  \hline
\end{tabular}
\end{table*}

\section{Results}
\label{sec:results}
\subsection{Throughout the years} 
\begin{figure}[ht]  
  \centering
   \includegraphics[scale=0.3]{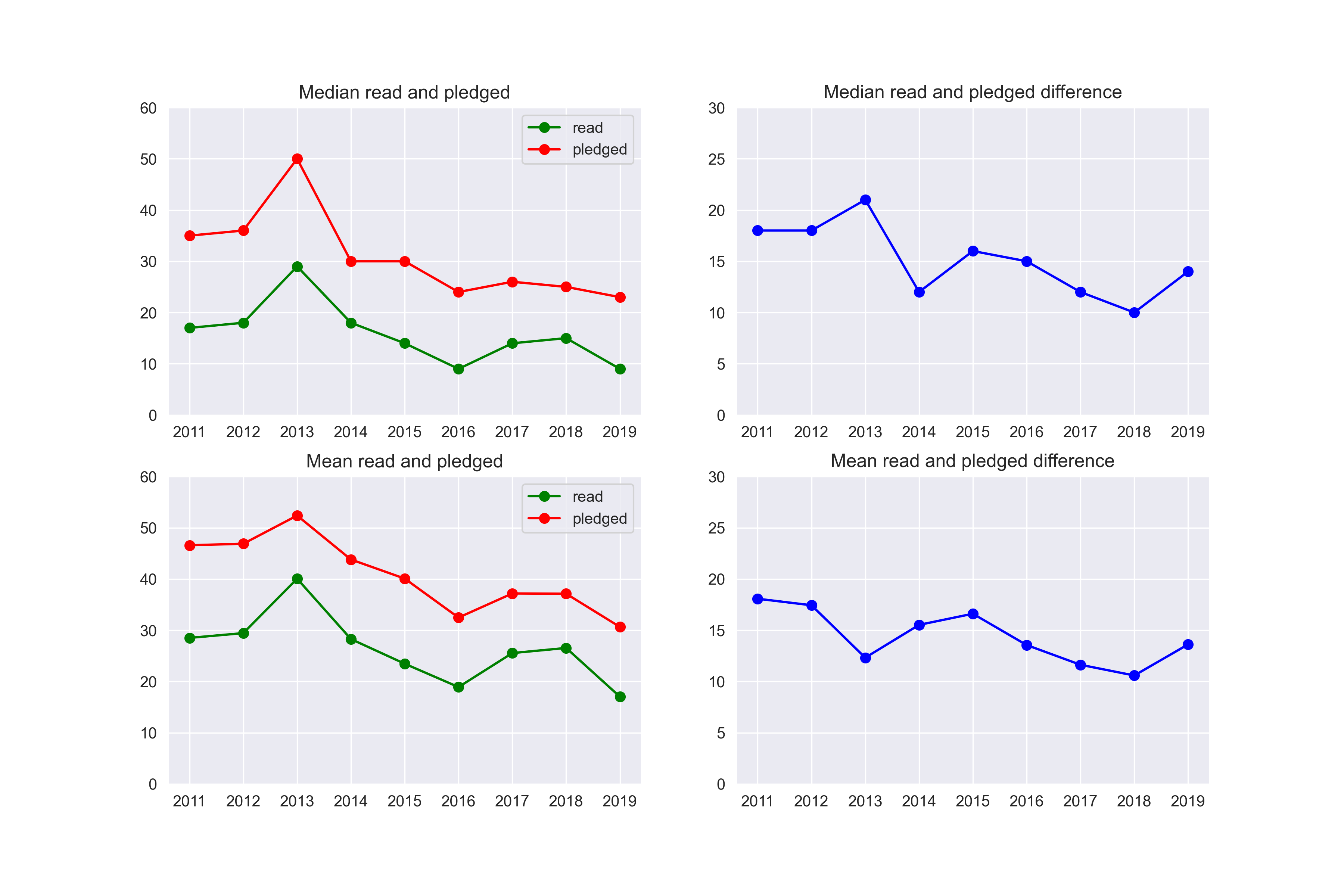}
  \captionsetup{justification=centering}
  \caption{The left-side plots show the median (top) and mean (bottom) count of pledged and read books for each year from 2011 to 2019. The right plots show the difference between the two for each year.}
  \label{fig:pledged_read_years}
\end{figure}

Our goal is to find how reading habits have changed through the years and whether people read less or more compared to the past. For this matter, we tried to find the trend in the read and pledged counts during challenges. By studying the average count of pledged books and read books, we realized that the average pledged count in challenges is 36.59 per challenge, and since each of these challenges correspond to a year, it means finishing 36.59 books every year, or 3.04 books every month. Conversely, the average number of books read during challenges is 23.30, which means finishing 1.94 books per month. This difference shows that people tend to read less but aim higher. It is crucial to bear in mind that this number of read books is slightly higher in a reading community and cannot be generalized to the entire society. For instance, based on a research \cite{site:average_number_of_read_books}, a person reads 12 books a year on average, or in other words, one book a month.\\ 
We also wanted to see how this overestimation has changed throughout the years. In order to find out, we grouped the challenge data by their year and calculated the median and mean pledged and read count for each year separately. The difference between these two counts can be seen in Figure \ref{fig:pledged_read_years}. 
Based on these figures, the difference between the pledged count and read count has mainly decreased, and people tend to better estimate their abilities compared to the past. The reason to plot both the mean and the median is that the median is less sensitive to outliers. It is also apparent that people are generally reading and pledging less than before.

\subsection{Do challenges make people read more?}
In this part, the data collected from users' profiles containing all the challenges they had participated in was used. Each book that these users had read during different years was also retrieved and collected in a separate dataset. This way, we could find out how many books these users had read through the years and also had the information about which year they were participating in an annual challenge. After calculating the average number of books each user reads in years that they are not participating in a yearly challenge and the years they are and comparing them, it was concluded that 81\% of people read more on average while being part of a reading challenge. This information belongs to the users that we have information on their readings both in years they were participating in challenges and years they were not, namely 787 unique users. Users read 298\% more books on average while participating in a challenge than when they are not. In order to find out whether this difference is significant, we performed a hypothesis test: $H_0: \mu_1 - \mu_2 = 0$ and $H_A: \mu_1 - \mu_2 \ne 0$, where $\mu_1$ and $\mu_2$ are the average read books during challenges and outside them, respectively. 

\begin{figure}[ht]  
  \centering
   \includegraphics[scale=0.25]{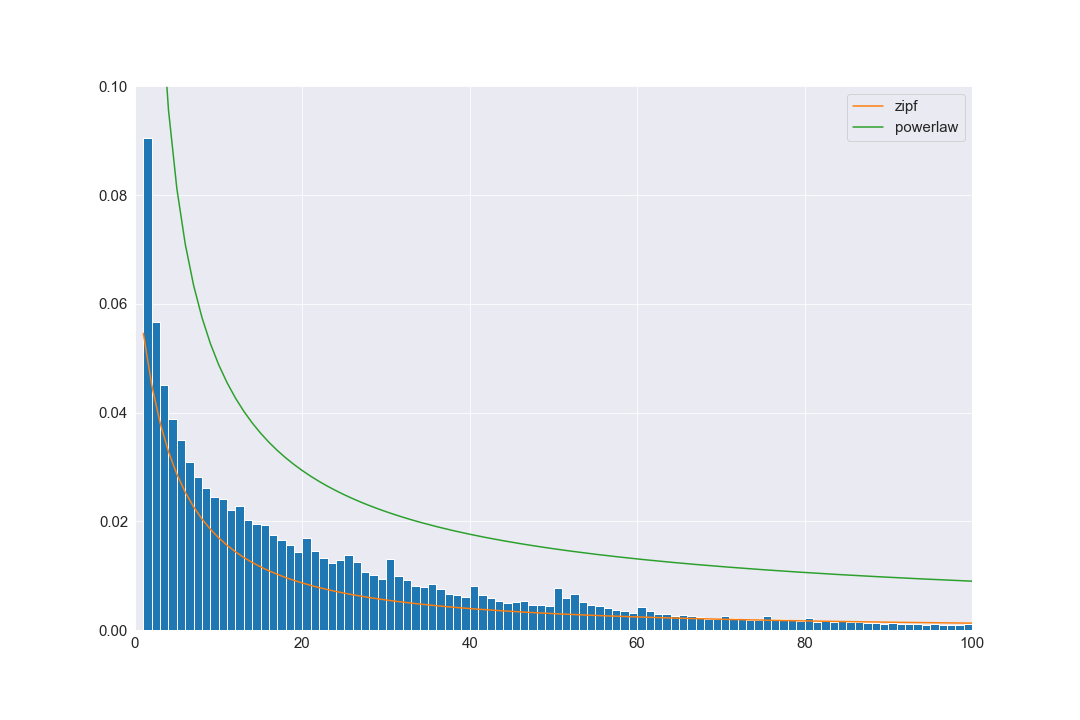}
  \captionsetup{justification=centering}
  \caption{Powerlaw vs Zipf Distributions for the Read Count Histogram}
  \label{fig:zipf_vs_powerlaw}
\end{figure}

This test's p-value is almost 0, which shows that the observed difference is statistically significant, and people read more books while taking part in yearly challenges. The reason could be that users feel more motivated to reach their goal when publicly announcing it or are more likely to update their Goodreads profile and their reading progress while participating in a challenge.
 \\
 Another interesting result is the number of users who have read a specific number of books in a challenge. This data is depicted in Figure \ref{fig:zipf_vs_powerlaw}. Before plotting this histogram, we assumed that it would have a normal distribution as it corresponds to human performance. However, the result showed a completely different distribution. We tried to fit power law and Zipf distributions, and they had the sum of squared errors 0.070 and 0.0027, respectively.
 \\
 
\subsection{Are demographic variables influential in reading habits?}
This section investigates whether people's demographic identities, including gender and place of residence, significantly influence their reading habits and preferences. \\
The questions below were studied in this section:
\begin{itemize}
    \item What are the reading challenge success rates in different countries?
    \item Is gender statistically significant in whether people read audiobooks or not?
    \item Is gender statistically significant in succeeding at Goodreads reading challenges?
\end{itemize}

\begin{figure}[ht]  \centering
   \includegraphics[scale=0.255]{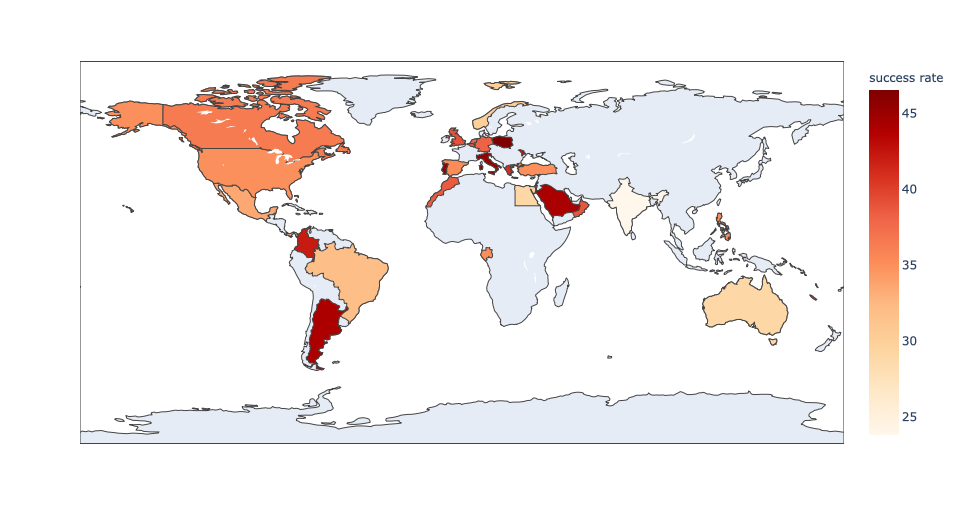}
  \captionsetup{justification=centering}
  \caption{Success Rate in Countries}
  \label{fig:success_countries}
\end{figure}

\begin{figure}[ht]  \centering
   \includegraphics[scale=0.255]{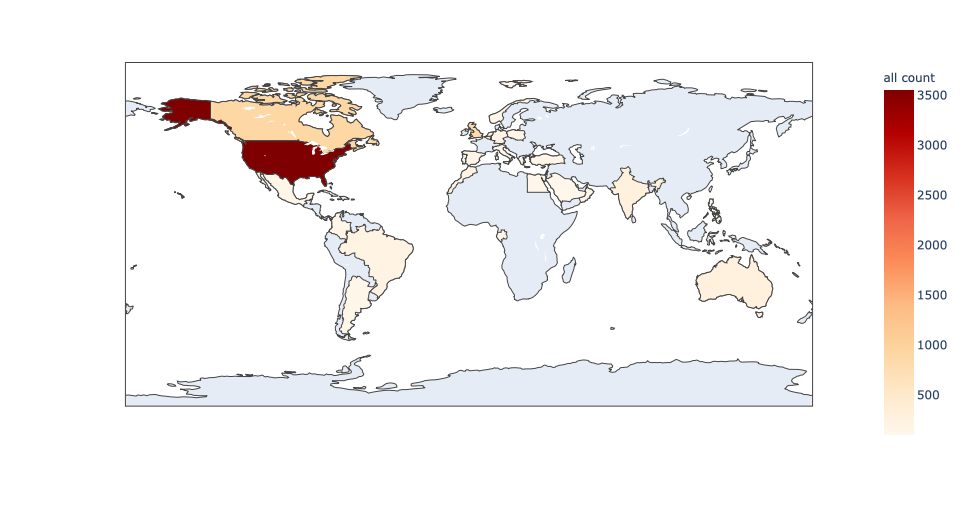}
  \captionsetup{justification=centering}
  \caption{Challenge Count in Countries}
  \label{fig:count_countries}
\end{figure}

After extracting the country names for users based on the location and personal details, each country's success rate was calculated. To avoid bias, the countries with less than 100 challenges were not considered in this calculation. After the filtering, 27 countries were left. The success rate for these countries is shown in Figure~\ref{fig:success_countries}. The most successful countries were Poland, Portugal, and Italy, with success rates of 0.4656, 0.4552, and 0.4522. Furthermore, the number of challenges in each of these countries is shown in Figure \ref{fig:count_countries}. This figure shows how much of our data belonged to each country. The top three countries are the United States, Canada, and the United Kingdom, with 3557, 888, and 790 challenges, respectively. \\

Another question that we will address here is whether gender has a significant role in users' tendency to read audiobooks during challenges. To find each user's gender belonging to the dataset containing each user and their books, \textit{gender-guesser}\footnote{https://pypi.org/project/gender-guesser/} library was used; this dataset was created by performing inner join on the challenge dataset and the books they had read during these challenges. There were 124,263 male and 575,406 female users in this data, and 303,245 users' gender could not be detected. Proportion difference hypothesis test is used to investigate whether gender has a significant role in choosing audiobooks over other book formats during challenges. (See Table \ref{tab:audiobook_gender}). 

Based on the results of a hypothesis test we performed where $H_0: p_{male} - p_{female} = 0$ and $H_1: p_{male} - p_{female} > 0$, the p-value for one-tailed hypothesis is $<$ .00001 and the result is significant at the significance level 0.05. As a result, We can conclude that male users are more likely to choose audiobooks over other forms of books during challenges.\\
The other question was whether gender is significant in reading challenge success rate (See Table \ref{tab:success_gender}).
Based on a hypothesis test we performed where $H_0: p_{female} - p_{male} = 0$ and $H_1: p_{female} - p_{male} > 0$, the p-value is $<$ .00001 and the result is significant at the significance level 0.05. To be more specific, considering the count of challenge participants of each gender and their success rate, it can be concluded that gender has a significant influence on Goodreads users' success in annual reading challenges, and women are more successful in these challenges.\\ \\

\begin{table}[htbp]
  \centering
  \captionsetup{justification=centering}
  \caption{People Reading Audiobooks}
  \label{tab:audiobook_gender}
  \begin{tabular}{c c c}
    \hline
    Gender&People reading audiobooks count&All count\\
    \hline
    Male & 1731 & 124263  \\ \hline
    Female & 6424 & 575406 \\
  \hline
\end{tabular}
\end{table}

\begin{table}[htbp]
  \centering
  \captionsetup{justification=centering}
  \caption{People Success Count}
  \label{tab:success_gender}
  \begin{tabular}{c c c}
    \hline
    Gender&Successful count&All count\\
    \hline
    Male & 154,010 & 533,153  \\ \hline
    Female & 424,566 & 1,394,773 \\
  \hline
\end{tabular}
\end{table}

\subsection{Other social networks}
\begin{figure}[ht]  
  \centering
  \captionsetup{justification=centering}
   \includegraphics[scale=0.25]{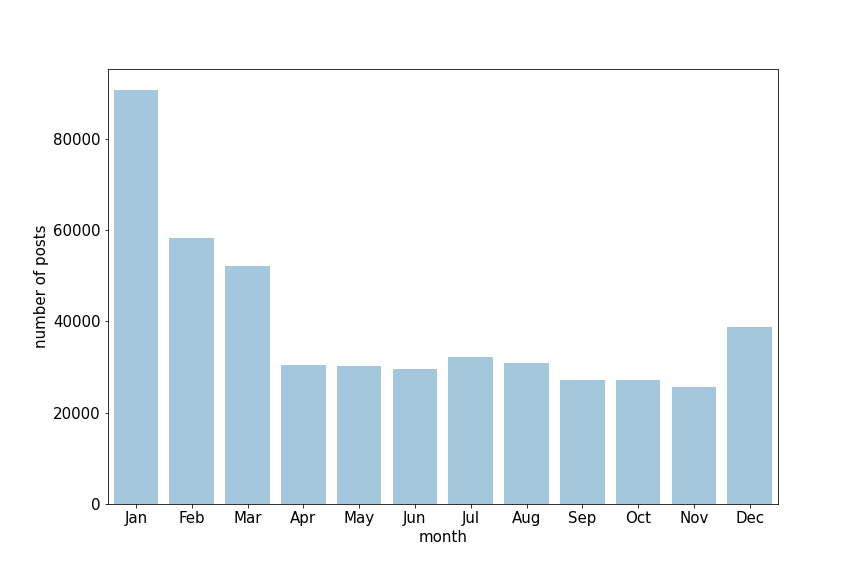}
  \caption{Instagram Posts Count through Months}
  \label{fig:insta_count_months}
\end{figure}

\begin{figure}[ht]  
  \centering
  \captionsetup{justification=centering}
   \includegraphics[scale=0.25]{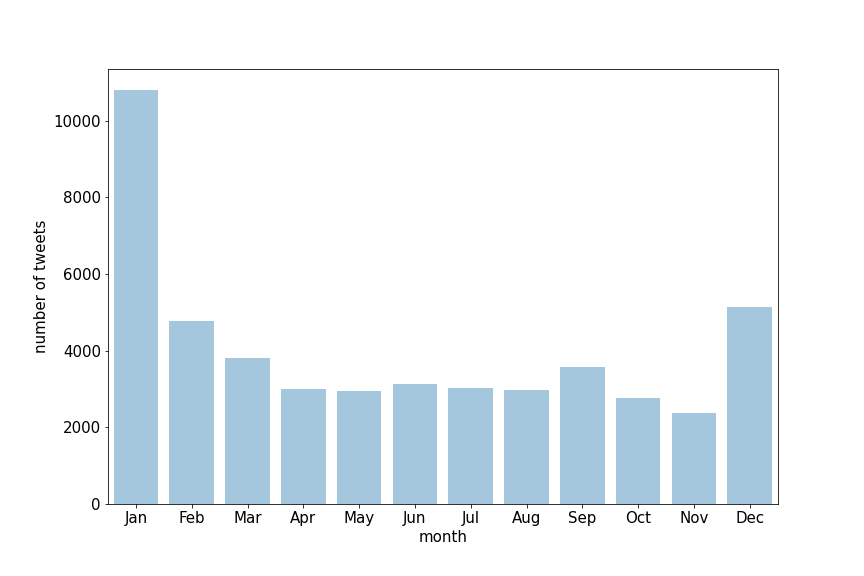}
  \caption{Tweets Count through Months}
  \label{fig:twitter_count_months}
\end{figure}

We analyzed the number of reading challenge tweets and posts through different months of the year. Generally, people announce their goals at the beginning of the year as these reading challenges are annual events, and people usually set goals as their new year's resolution. For instance, 19\% of Instagram posts and 22\% of tweets regarding reading challenges were posted in January. They will keep posting less and less in the following months until December when the challenge is close to the end, and a sudden increase in the number of posts is observed then. The number of reading challenge Instagram posts and tweets through months are depicted in Figure \ref{fig:insta_count_months} and Figure \ref{fig:twitter_count_months}, respectively, which confirm the information given above.

\begin{figure}[ht]  
  \centering
  \captionsetup{justification=centering}
   \includegraphics[scale=0.25]{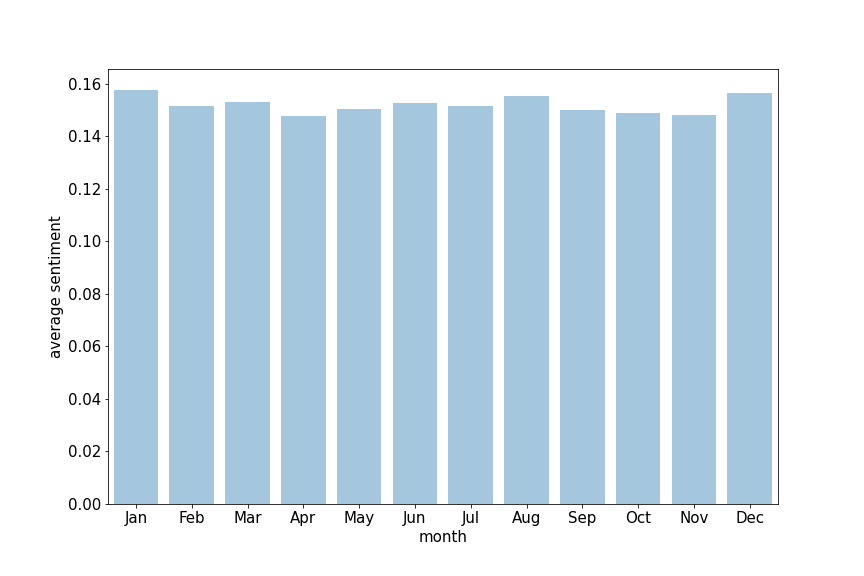}
  \caption{Average Sentiment of Instagram Posts through Months}
  \label{fig:insta_sentiment_months}
\end{figure}

\begin{figure}[ht]  
  \centering
  \captionsetup{justification=centering}
   \includegraphics[scale=0.25]{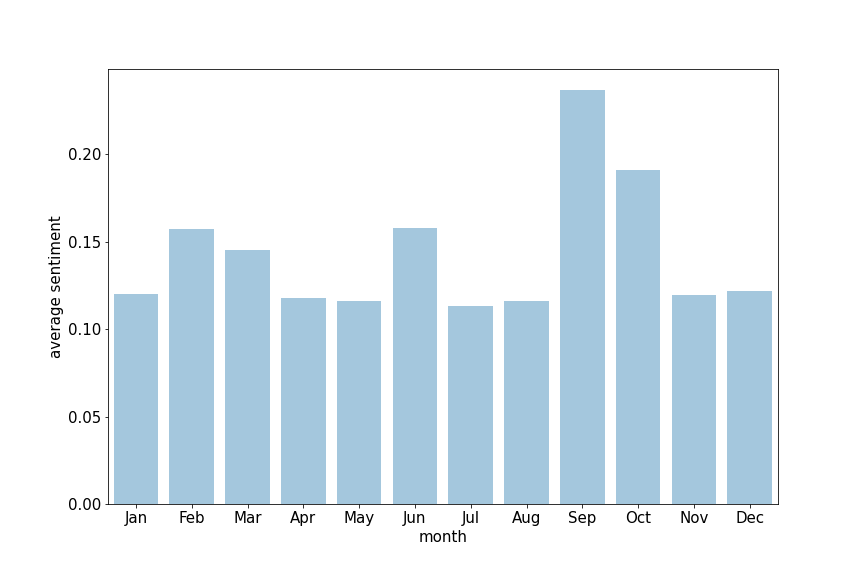}
  \caption{Average Sentiment of Tweets through Months}
  \label{fig:twitter_sentiment_months}
\end{figure}

The sentiment analysis results show that users' posts regarding reading challenges are rarely negative. Users might merely post a report on their progress, express their feeling about a specific book, or how successful they are in their challenge. They might also express their disappointment in falling behind during the challenge. Some users also post reviews about books and add their opinions about them, which can be positive or negative. The average sentiment through months in Instagram and Twitter are shown in Figure \ref{fig:insta_sentiment_months} and Figure \ref{fig:twitter_sentiment_months}, respectively. In Instagram, the general attitude towards reading challenges is similar through months, but Twitter data shows some fluctuations in the tweets' sentiments. However, the difference between the highest and lowest average sentiments in months is merely 0.123, and therefore, we can assume that the general attitude is similar.

The percentage of posts with negative and positive sentiments on both social platforms are almost identical. This is depicted in Table \ref{tab:sentiment}. This shows that people on both platforms have similar attitudes toward their reading challenges.

\begin{table}[htbp]
  \centering
  \captionsetup{justification=centering}
  \caption{Sentiment Percentage}
  \label{tab:sentiment}
  \begin{tabular}{c c c c}
    \hline
    Platform&Positive&Negative&Neutral\\
    \hline
    Instagram &50.54\% & 4.77\%&44.67\% \\ \hline
    Twitter & 47.74\% & 4.64\%&47.60\% \\
  \hline
\end{tabular}
\end{table}

\section{Conclusion}
\label{sec:conclusion}
In this study, we examined the effects of public reading challenges on how much people read. Specifically, we showed that people are significantly more likely to read more once they have taken part in a challenge in comparison to their normal performance. We further show how gender is a significant factor in how well people perform in their challenges. \\
We would like to discuss one limitation of our study now. This analysis was conducted on Goodreads, and while the selection process was conducted completely at random, the website itself tends to attract avid readers mostly. Hence, the sample is out of people with a keen interest in reading, introducing bias into our analysis. 

\section{Declarations}
\subsection{Funding}
Not applicable / No funding was received.

\subsection{Conflicts of interest}
The authors declare that they have no competing interests.

\subsection{Availability of data and material}
 Although all data was collected from public accounts, we have decided not to make this data publicly available due to the availability of user information in our data. 
 
\subsection{Code availability}
Not applicable

\bibliographystyle{spmpsci}
\bibliography{goodreads}

\end{document}